\documentclass[preprint,showpacs,preprintnumbers,amsmath,amssymb]{revtex4}
\usepackage{epsfig}
\usepackage{graphicx}
\usepackage{dcolumn}
\usepackage{bm}
\usepackage{threeparttable}

\def\beq{\begin{equation}}
\def\eeq{\end{equation}}
\def\eeqn{\end{equation}}
\newcommand\iden{\leavevmode\hbox{\small1\normalsize\kern-.33em1}}

\newcommand{\bea} {\begin{eqnarray}}
\newcommand{\eea} {\end{eqnarray}}

\let\jnfont=\rm
\def\NPB#1,{{\jnfont Nucl.\ Phys.\ B }{\bf #1},}
\def\PLB#1,{{\jnfont Phys.\ Lett.\ B }{\bf #1},}
\def\EPJC#1,{{\jnfont Eur.\ Phys.\ Jour.\ C }{\bf #1},}
\def\PRD#1,{{\jnfont Phys.\ Rev.\ D }{\bf #1},}
\def\PRL#1,{{\jnfont Phys.\ Rev.\ Lett.\ }{\bf #1},}
\def\MPLA#1,{{\jnfont Mod.\ Phys.\ Lett.\ A }{\bf #1},}
\def\JPG#1,{{\jnfont J.\ Phys.\ G }{\bf #1},}
\def\CTP#1,{{\jnfont Commun.\ Theor.\ Phys.\ }{\bf #1},}
\def\JHEP#1,{{\jnfont JHEP \ }{\bf #1},}
\def\NPPS#1,{{\jnfont Nucl.\ Phys.\ Proc.\ Suppl.\ }{\bf #1},}
\def\CPC#1,{{\jnfont Comput.\ Phys.\ Commun.\ }{\bf #1},}
\def\CPL#1,{{\jnfont Chin.\ Phys.\ Lett. }{\bf #1},}
\def\APPB#1,{{\jnfont Acta\ Phys.\ Polon.\ B }{\bf #1},}
\def\PR#1,{{\jnfont Phys.\ Rept.\  }{\bf #1},}
\def\CHC#1,{{\jnfont Chin.\ Phys.\ C }{\bf #1},}

\def\lsim{\raise0.3ex\hbox{$<$\kern-0.75em\raise-1.1ex\hbox{$\sim$}}}
\def\gsim{\raise0.3ex\hbox{$>$\kern-0.75em\raise-1.1ex\hbox{$\sim$}}}

\begin{document}

\title{\ \\[10mm] Higgs pair signal enhanced in the 2HDM with two degenerate\\
125 GeV Higgs bosons}

\author{Xiao-Fang Han$^{1}$, Lei Wang$^{2,1}$, Jin Min Yang$^{3}$}
 \affiliation{$^1$ Department of Physics, Yantai University, Yantai
264005, P. R. China\\
$^2$ IFIC, Universitat de Val$\grave{e}$ncia-CSIC, Apt. Correus
22085, E-46071 Val$\grave{e}$ncia, Spain\\
$^3$ State Key Laboratory of Theoretical Physics,
      Institute of Theoretical Physics, Academia Sinica, Beijing 100190,
      P. R. China}

\begin{abstract}
We discuss a scenario of the type-II 2HDM in which the
$b\bar{b}\gamma\gamma$ rate of the Higgs pair production is enhanced
due to the two nearly degenerate 125 GeV Higgs bosons ($h$, $H$).
Considering various theoretical and experimental constraints, we
figure out the allowed ranges of the trilinear couplings of these
two Higgs bosons and calculate the signal rate of
$b\bar{b}\gamma\gamma$ from the productions of Higgs pairs ($hh$,
$hH$, $HH$) at the LHC. We find that in the allowed parameter space
some trilinear Higgs couplings can be larger than the SM value by an
order and the production rate of $b\bar{b}\gamma\gamma$ can be
greatly enhanced. We also consider a "decoupling" benchmark point
where the light CP-even Higgs has a SM-like cubic self-coupling
while other trilinear couplings are very small. With a detailed
simulation on the $b\bar{b}\gamma\gamma$ signal and backgrounds, we
find that in such a "decoupling" scenario the $hh$ and $hH$ channels
can jointly enhance the statistical significance to 5$\sigma$ at 14
TeV LHC with an integrated luminosity of 3000 fb$^{-1}$.

\end{abstract}
 \maketitle

\section{Introduction}
So far the properties of the 125 GeV Higgs boson discovered by the
ATLAS and CMS collaborations \cite{cmsh,atlh} agree with the
Standard Model (SM) predictions. However, there is no experimental
information for the Higgs self-coupling, which is vital for the
spontaneous electroweak symmetry breaking. As is well known, the
Higgs pair production at the LHC may provide a way to probe the
Higgs self-coupling. The signal $b\bar{b}b\bar{b}$ from the Higgs
pair has the largest rate, but suffers from the huge QCD background.
The $bb\tau\bar{\tau}$ channel is swamped by the $b\bar{b}jj$
background \cite{bbtautau} where each light-flavored jet can fake a
hadronic $\tau$. The detection of these two channels and also the
$b\bar{b}WW^*$ channel needs more elaborated strategies like boosted
kinematics and jet substructure technique \cite{subjet}. Although
the $b\bar{b}\gamma\gamma$ channel has a small rate, it has the
cleanest background, and thus has attracted more attention
\cite{yao,1502.00539,caohh,hh2h1,hh2h2,100tev}. For the SM, the
significance for $gg \to hh \to b\bar{b} \gamma\gamma$ is only
around 2$\sigma$ at the 14 TeV LHC with an integrated luminosity of
3000 fb$^{-1}$ \cite{yao,1502.00539}. So a collider with higher
energy (say 100 TeV) seems needed to examine the Higgs self-coupling
from the Higgs pair production.

The Higgs pair production can serve as a good probe for new physics.
The production rate can be enhanced by modifying the Higgs
self-coupling or top quark Yukawa coupling properly. Also, it can be
enhanced by some new mechanisms in the production, such as the heavy
top partner loops in the little Higgs model \cite{hhlh}, the squark
loops in the SUSY models \cite{hhsusy}, and the on-shell production
of a heavy Higgs which decays into a pair of 125 GeV Higgses in the
two-Higgs-doublet model (2HDM) \cite{hh2h1,hh2h2}. In this work, we
will discuss a scenario in the type-II 2HDM  \cite{type-ii} where
the $b\bar{b}\gamma\gamma$ channel of the Higgs pair is enhanced due
to the two nearly degenerate 125 GeV Higgses (similar degenerate
cases have been discussed in the literature \cite{gunion}, but their
impact on Higgs pair signals has not been studied). The mass
splitting between these two Higgses is smaller than the mass
resolution of the detector while larger than their widths so that
the interference terms can be neglected. First, considering the
theoretical constraints from vacuum stability, unitarity and
perturbativity as well as the experimental constraints from the
electroweak precision data, flavor observables and Higgs data, we
will figure out the allowed ranges of the trilinear couplings of
these two Higgses and calculate the $b\bar{b}\gamma\gamma$
production rate at the LHC. Then, focusing on a "decoupling"
benchmark point where the light CP-even Higgs has a SM-like cubic
self-coupling while the other trilinear couplings are very small, we
perform a detailed simulation on the $b\bar{b}\gamma\gamma$ signal
and its backgrounds at the 14 TeV LHC with an integrated luminosity
of 3000 fb$^{-1}$ .

Our work is organized as follows. In Sec. II we recapitulate the
type-II 2HDM. In Sec. III we describe our numerical calculations.
In Sec. IV, we show the allowed ranges of the various trilinear couplings
of the two Higgses and give the simulation results for the $b\bar{b}\gamma\gamma$ signal
and its backgrounds at the LHC.
Finally, we draw our conclusion in Sec.V.

\section{Type-II 2HDM}
The general Higgs potential of 2HDM is written as \cite{2h-poten}
\begin{eqnarray} \label{V2HDM} \mathrm{V} &=& m_{11}^2
(\Phi_1^{\dagger} \Phi_1) + m_{22}^2 (\Phi_2^{\dagger}
\Phi_2) - \left[m_{12}^2 (\Phi_1^{\dagger} \Phi_2 + \rm h.c.)\right]\nonumber \\
&&+ \frac{\lambda_1}{2}  (\Phi_1^{\dagger} \Phi_1)^2 +
\frac{\lambda_2}{2} (\Phi_2^{\dagger} \Phi_2)^2 + \lambda_3
(\Phi_1^{\dagger} \Phi_1)(\Phi_2^{\dagger} \Phi_2) + \lambda_4
(\Phi_1^{\dagger}
\Phi_2)(\Phi_2^{\dagger} \Phi_1) \nonumber \\
&&+ \left[\frac{\lambda_5}{2} (\Phi_1^{\dagger} \Phi_2)^2 + \rm
h.c.\right] + \left[\lambda_6 (\Phi_1^{\dagger} \Phi_1)
(\Phi_1^{\dagger} \Phi_2) + \rm h.c.\right] \nonumber \\
&& + \left[\lambda_7 (\Phi_2^{\dagger} \Phi_2) (\Phi_1^{\dagger}
\Phi_2) + \rm h.c.\right].
\end{eqnarray}
In the type-II 2HDM, a discrete $Z_2$ symmetry is introduced to make
$\lambda_6=\lambda_7=0$ while allow for a soft-breaking term with
$m_{12}^2\neq 0$. All $\lambda_i$ and $m_{12}^2$ are taken to be
real in order to avoid the explicit CP violation in the Higgs
sector.

The two complex scalar doublets have the hypercharge $Y = 1$,
\begin{equation}
\Phi_1=\left(\begin{array}{c} \phi_1^+ \\
\frac{1}{\sqrt{2}}\,(v_1+\phi_1^0+ia_1)
\end{array}\right)\,, \ \ \
\Phi_2=\left(\begin{array}{c} \phi_2^+ \\
\frac{1}{\sqrt{2}}\,(v_2+\phi_2^0+ia_2)
\end{array}\right),
\end{equation}
where $v_1$ and $v_2$ are the vacuum expectation values (VEVs) with
$v^2 = v^2_1 + v^2_2 = (246~\rm GeV)^2$ and $\tan\beta$ is
defined as $v_2 /v_1$. The physical scalar spectrum of this model
consists of two neutral CP-even $h$ and $H$, one neutral
pseudoscalar $A$, and two charged scalar $H^{\pm}$. This basis can
be rotated to the Higgs basis by a mixing angle $\beta$, where the
VEV of $\Phi_2$ field is zero. In the Higgs basis, the mass
eigenstates are obtained from
\bea
&&h=\sin(\beta-\alpha)\phi^0_1+\cos(\beta-\alpha)\phi^0_2, \nonumber\\
&&H=\cos(\beta-\alpha)\phi^0_1-\sin(\beta-\alpha)\phi^0_2,
\nonumber\\
&&A=a_2,~~~~~~~~~~H^\pm=\phi^\pm_2,
\eea
where the fields in the righ sides denote the interaction eigenstates in the Higgs basis.

In the Higgs basis, the general Yukawa interactions with no
tree-level FCNC are written as \cite{a2hm-1}
\begin{equation} \label{eq:Yukawa1}
 \mathcal{L}_Y = -\frac{\sqrt{2}}{v}\,\Big[M'_d\bar{Q}_L ( \Phi_1 + \kappa_d \Phi_2) d_R
 +M'_u  \bar{Q}_L (\tilde{\Phi}_1 + \kappa_u \tilde{\Phi}_2) u_R
 + M'_\ell\bar{L}_L ( \Phi_1 + \kappa_\ell \Phi_2) \ell_R \Big]
  + \mathrm{h.c.} \,,
\end{equation}
where $\tilde{\Phi}_i(x)=i\tau_2\Phi_i^{\ast}(x)$ and
$M'_{d,u,\ell}$ are the Yukawa matrices.  For the type-II 2HDM, we have
\beq
\kappa_u=\cot\beta,~~~\kappa_d=\kappa_\ell=-\tan\beta.
\eeq
From Eq. (\ref{eq:Yukawa1}) we can obtain the couplings of neutral Higgs
bosons normalized to the SM Higgs boson
\bea &&
y^h_{V}=\sin(\beta-\alpha),~~~y^h_f=\sin(\beta-\alpha)+\cos(\beta-\alpha)\kappa_f,\nonumber\\
&&y^H_{V}=\cos(\beta-\alpha),~~~y^H_f=\cos(\beta-\alpha)-\sin(\beta-\alpha)\kappa_f,\nonumber\\
&&y^A_{V}=0,~~~~~~~y^A_u=-i\gamma^5
\kappa_{u},~~~~~~~~y^A_{d,\ell}=i\gamma^5
\kappa_{d,\ell},\label{yukawatop}
\eea
where $V$ denotes $Z$ and $W$,
and $f$ denotes $u$, $d$ and $\ell$. The charged Higgs couplings are
written as
\begin{align} \label{eq:Yukawa2}
 \mathcal{L}_Y & = - \frac{\sqrt{2}}{v}\, H^+\, \Big\{\bar{u} \left[\kappa_d\,V_{CKM} M_d P_R
 - \kappa_u\,M_u V_{CKM} P_L\right] d + \varsigma_\ell\,\bar{\nu} M_\ell P_R \ell
 \Big\}+h.c.,
 \end{align}
where $M_f$ are the diagonal fermion mass matrices.

\section{numerical calculations}
We employ $\textsf{2HDMC-1.6.5}$ \cite{2hc-1} to implement the
theoretical constraints from the vacuum stability, unitarity and
coupling-constant perturbativity, and calculate the oblique
parameters ($S$, $T$, $U$) and $\delta\rho$.
We use $\textsf{SuperIso-3.4}$ \cite{spriso} to implement the constraints
from  $B\to X_s\gamma$ and use $\textsf{HiggsBounds-4.1.3}$ \cite{hb} to
implement the exclusion constraints from the neutral and charged
Higgses searches at the LEP, Tevatron and LHC at 95\% confidence
level. The in-house code is used to calculate $\chi^2$ fit to 125.5
GeV Higgs signal, $\Delta m_{B_s}$
 and $\Delta m_{B_d}$.  In addition to the theoretical
constraints, we require the type-II 2HDM to satisfy
all the experimental data at 2$\sigma$ level. The
experimental values of electroweak precision data, $B\to X_s\gamma$,
$\Delta m_{B_s}$ and $\Delta m_{B_d}$ are taken from \cite{pdg2014}.

We generate the $\textsf{2HDM@NLO}$ model using the tree-level 2HDM
model and $\textsf{NLOCT}$ package \cite{nloct}. The model contains
the QCD R2 vertice and UV counterterms for the 2HDM, which is based
on the $\textsf{FeynRules}$ \cite{fr} and $\textsf{UFO}$ \cite{ufo}
frameworks. In our simulation, the parton level signal and
background events are generated with
$\textsf{MadGraph5$_{-}$aMC$_{-}$v2.3.0}$ \cite{mg5}. For the Higgs
pair production via gluon-gluon fusion, we take the factorization
and renormalization scales as the the invariant mass of the Higgs
pair. The in-house code is used to transform the results of
$\textsf{2HDMC}$ into the parameter card which is read by
$\textsf{MadGraph5$_{-}$aMC$_{-}$v2.3.0}$ since there are different
basis and mixing angles in CP-even Higgs sector between
$\textsf{2HDM@NLO}$ model and $\textsf{2HDMC}$. $\textsf{PYTHIA}$
\cite{pythia} is employed to decay the Higgs bosons following the
decay table of parameter card, and perform parton shower and
hadronization. We perform the fast detector simulations and data
analysis with $\textsf{Delphes}$ \cite{delphes} and
$\textsf{Madanalysis5}$ \cite{ma5}. Jet reconstruction is done using
the anti-$k_T$ algorithm with a radius parameter of $R =0.5$. The
efficiency for $b$-tagging is taken as $70\%$. The efficiency of
photon tagging and the mis-tagging of QCD jets is assumed to the
default value as in $\textsf{Delphes}$.

Using the method in \cite{chi}, we perform a global fit to the
125.5 GeV Higgs data of 29 channels after ICHEP 2014
\cite{higgsdata}. Since we assume that the mass splitting of the two
CP-even Higgses is smaller than the mass resolution of detector, the
signal strength for a  channel is defined as
\beq
\mu_i=\sum_{\hat{H}=h,~H}\epsilon_{gg\hat{H}}^i
R_{gg\hat{H}}+\epsilon_{VBF\hat{H}}^i
R_{VBF\hat{H}}+\epsilon_{V\hat{H}}^i
R_{V\hat{H}}+\epsilon_{t\bar{t}\hat{H}}^i R_{t\bar{t}\hat{H}},
\eeq
where $R_{j}=(\sigma \times BR)_j/(\sigma\times BR)_j^{SM}$
with $j$ denoting the partonic process
$gg\hat{H},~VBF\hat{H},~V\hat{H},$ or $t\bar{t}\hat{H}$,
and $\epsilon_{j}^i$ denotes the assumed signal composition of the
partonic process $j$ \cite{kmdata}, which has the same value for
$h$ and $H$. For an uncorrelated observable $i$,
\beq
\chi^2_i=\frac{(\mu_i-\mu^{exp}_i)^2}{\sigma_i^2},
\eeq
where $\mu^{exp}_i$ and $\sigma_i$ denote the experimental central value
and uncertainty for the $i$-channel. The uncertainty asymmetry is
retained in our calculations. For the two correlated observables, we
take
\beq
\chi^2_{i,j}=\frac{1}{1-\rho^2}
\left[\frac{(\mu_i-\mu^{exp}_i)^2}{\sigma_i^2}+\frac{(\mu_j-\mu^{exp}_j)^2}{\sigma_j^2}
-2\rho\frac{(\mu_i-\mu^{exp}_i)}{\sigma_i}\frac{(\mu_j-\mu^{exp}_j)}{\sigma_j}\right],
\eeq
where $\rho$ is the correlation coefficient. We sum over $\chi^2$ in
the 29 channels, and pay particular attention to the surviving
samples with $\chi^2-\chi^2_{\rm min} \leq 6.18$, where $\chi^2_{\rm
min}$ denotes the minimum of $\chi^2$. These samples correspond to
the 95.4\% confidence level region in any two-dimension plane of
the model parameters when explaining the Higgs data (corresponding
to the $2\sigma$ range).

In our calculations, we take $m_h=$ 125.5 GeV and $m_H=$ 126 GeV,
and the input parameters are $\cos(\beta-\alpha)$,
$\tan\beta$, the physical Higgs masses ($m_A$, $m_{H^{\pm}}$) and
the soft breaking parameter $m_{12}^2$. Since the Higgs couplings
between the two CP-even Higgses are independent of $m_A$ and
$m_{H^{\pm}}$, we take $m_A = m_{H^{\pm}}$, which is favored
by the $\delta\rho$ and oblique parameters. We scan randomly the
parameters in the following ranges
\bea
&&0\leq\cos(\beta-\alpha)\leq0.1~~,1 \leq\tan\beta \leq
15,\nonumber\\ &&200~{\rm GeV}\leq m_A=m_{H^{\pm}} \leq 700~ {\rm
GeV},~~-(400~{\rm GeV})^2 \leq m_{12}^2 \leq (400~{\rm GeV})^2.
\eea
We take the convention $0\leq\cos(\beta-\alpha)\leq 1$ and
$-1\leq\sin(\beta-\alpha)\leq 1$. With $0\leq\cos(\beta-\alpha)\leq 0.1$,
the couplings between the light CP-even Higgs and the gauge bosons are
close to the SM predictions while the corresponding heavy Higgs couplings
are very small.

\section{results and discussions}
After imposing the above mentioned theoretical and experimental constraints,
we find the minimal value of $\chi^2$ is $\chi^2_{min}\simeq 18.08$,
which is slightly larger than the SM value (17.0). And the corresponding parameters
are
\bea&&
\sin(\beta-\alpha)\simeq0.99996,~\tan\beta\simeq3.094,~m_h=125.5
~{\rm{GeV}},~m_H\simeq126.0~{\rm{GeV}},\nonumber\\
&&m_A=448.88~{\rm{GeV}},~m_{H^\pm}=448.88~{\rm{GeV}},m_{12}^2=4615.4~\rm{GeV^2}.
\eea

\subsection{Higgs pair cross section and Higgs trilinear couplings}
We define $R_{b\bar{b}\gamma\gamma}$ as the $b\bar{b}\gamma\gamma$
signal event number of type-II 2HDM normalized to the SM prediction
\beq
R_{b\bar{b}\gamma\gamma}=\frac{\sum\limits\sigma(gg\to
\hat{H}\hat{H})\times Br(\hat{H}\hat{H}\to
b\bar{b}\gamma\gamma)}{\sigma(gg\to hh)_{SM}\times Br(hh\to
b\bar{b}\gamma\gamma)_{SM}},
\eeq
where $\hat{H}\hat{H}$ denotes $hh$, $hH$ or $HH$. In fact, the contributions
from $gg\to HH\to b\bar{b}\gamma\gamma$ can be neglected since $Br(H\to \gamma\gamma)$
is much smaller than the SM prediction for $0\leq\cos(\beta-\alpha)\leq 0.1$.

\begin{figure}[tb]
 \epsfig{file=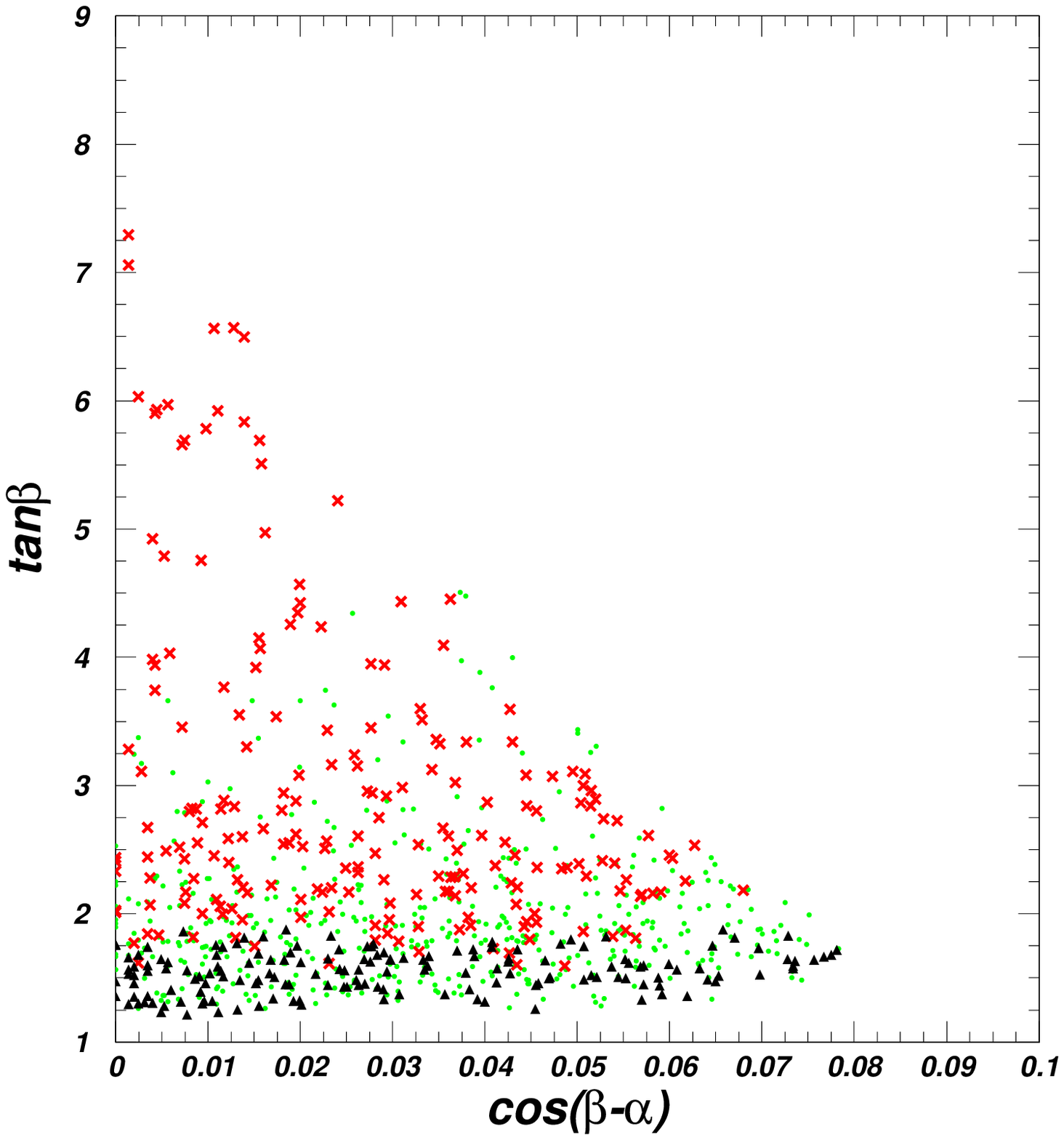,height=8.5cm}
 \epsfig{file=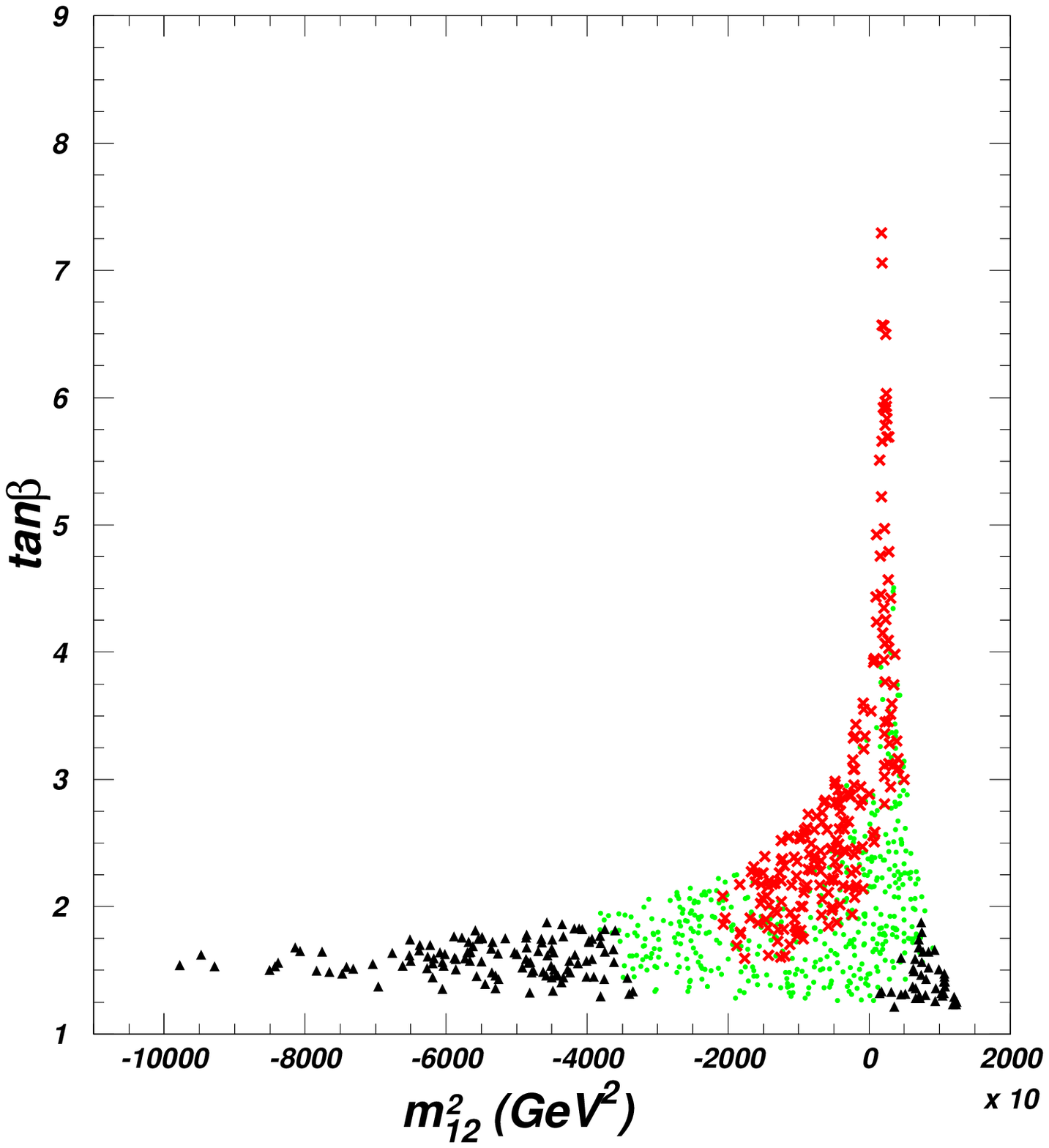,height=8.5cm}
\vspace{-0.3cm} \caption{The scatter plots of surviving samples
projected on the planes of $\cos(\beta-\alpha)$ versus $\tan\beta$
and $m_{12}^2$ versus $\tan\beta$, respectively. The crosses (red)
are for $R_{b\bar{b}\gamma\gamma}\leq1.2$, and bullets (green)
for $1.2<R_{b\bar{b}\gamma\gamma}\leq2.0$, and triangles (black) for
$R_{b\bar{b}\gamma\gamma}>2.0$.} \label{sbatb}
\end{figure}

\begin{figure}[tb]
 \epsfig{file=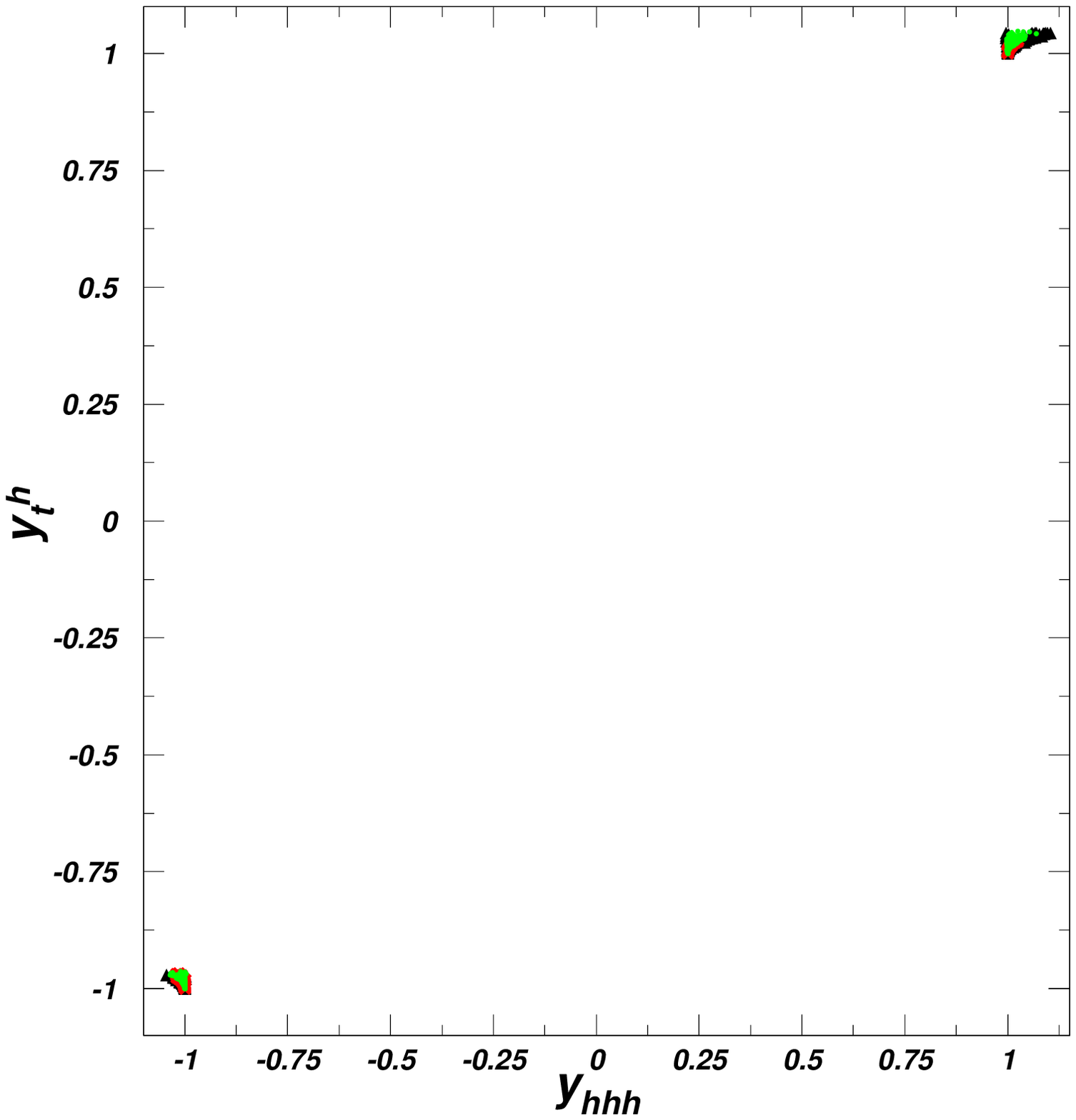,height=8.5cm}
 \epsfig{file=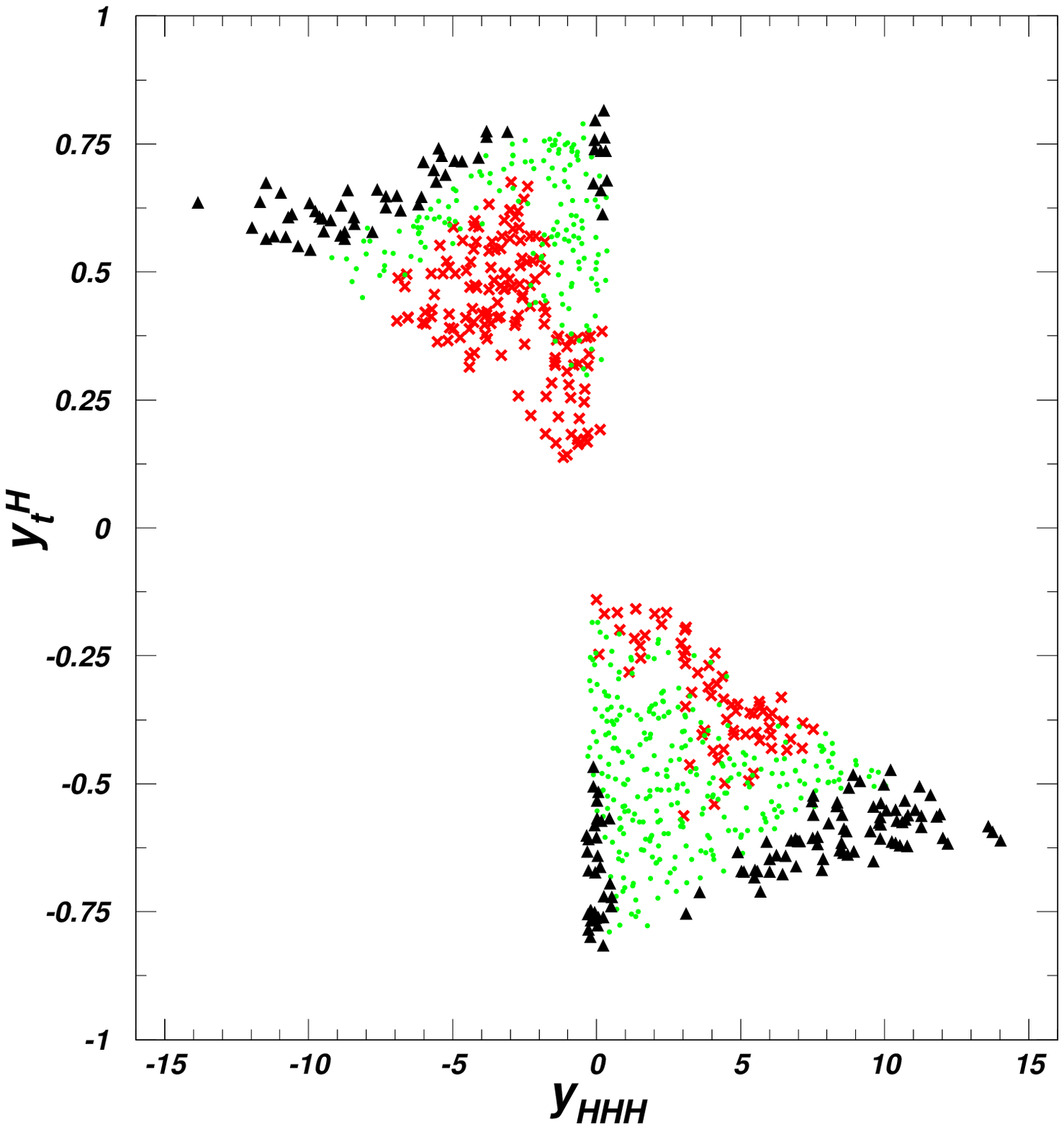,height=8.5cm}
  \epsfig{file=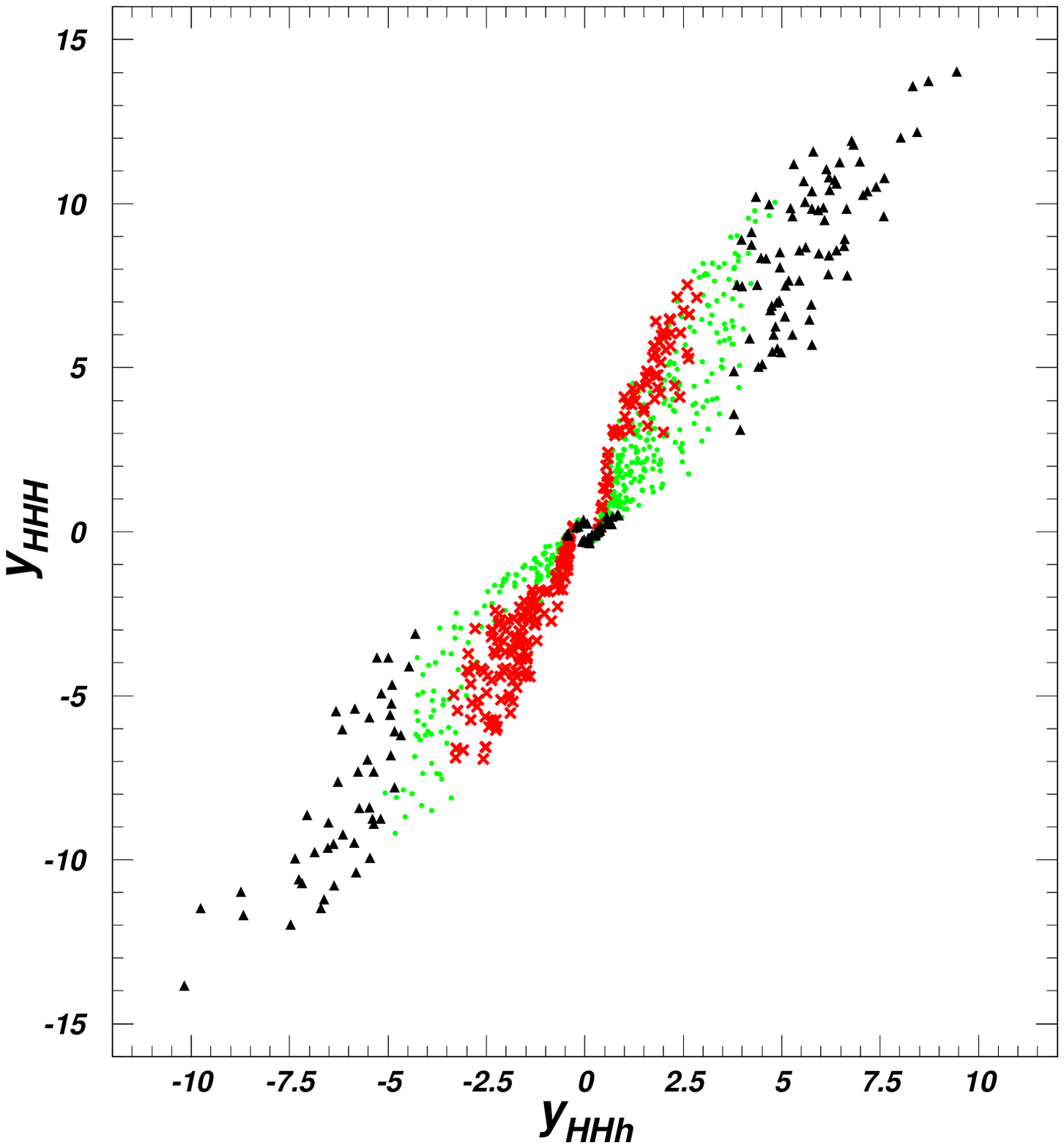,height=8.5cm}
 \epsfig{file=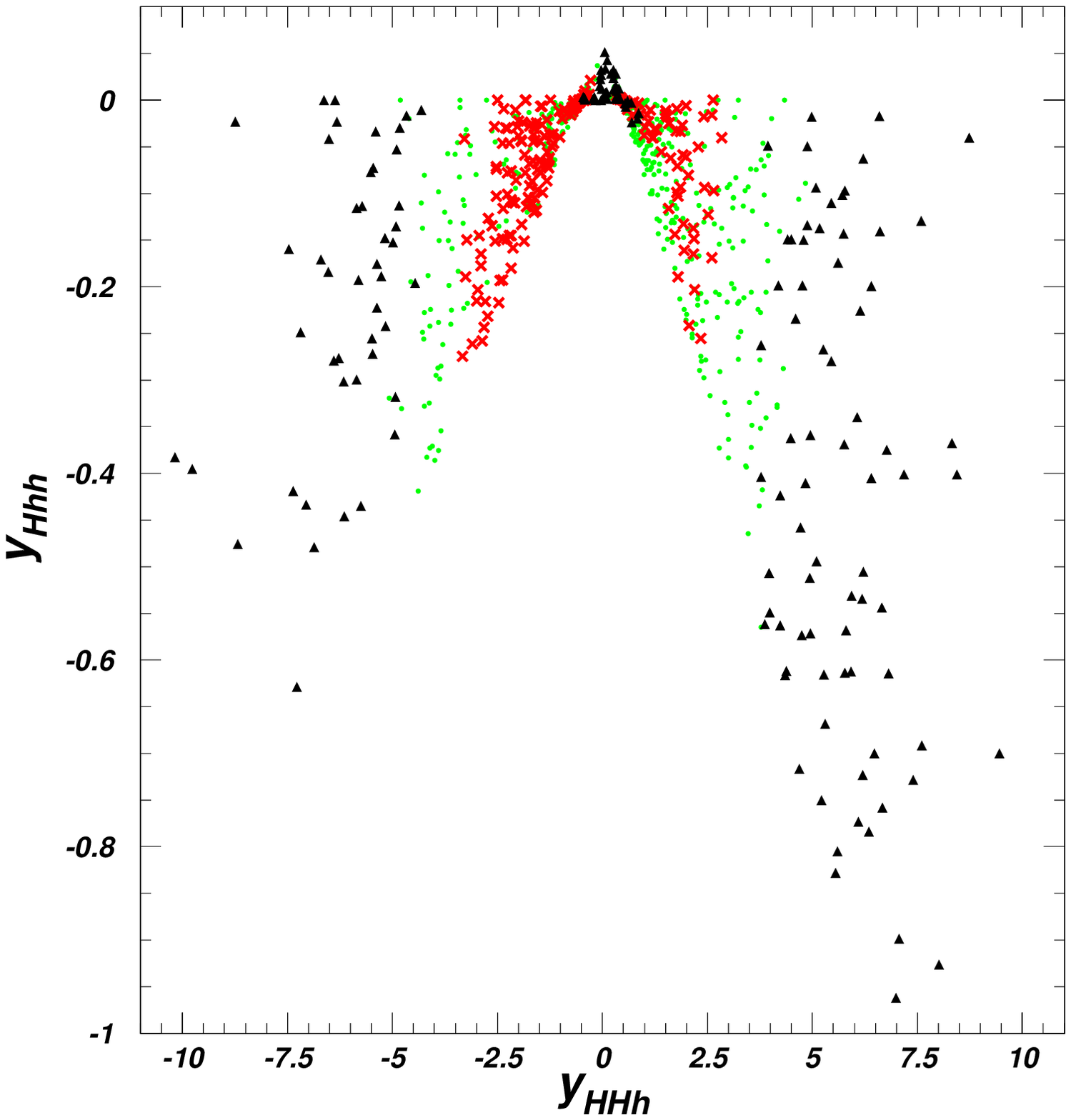,height=8.5cm}
\vspace{-0.3cm} \caption{Same as Fig. \ref{sbatb}, but projected on
the planes of $y_t^h$ versus $y_{hhh}$, $y_t^H$ versus $y_{HHH}$,
$y_{HHH}$ versus $y_{HHh}$ and $y_{Hhh}$ versus $y_{HHh}$.
All these Higgs trilinear couplings are
normalized to the SM $hhh$ coupling.} \label{rhcoup}
\end{figure}

\begin{figure}[tb]
 \epsfig{file=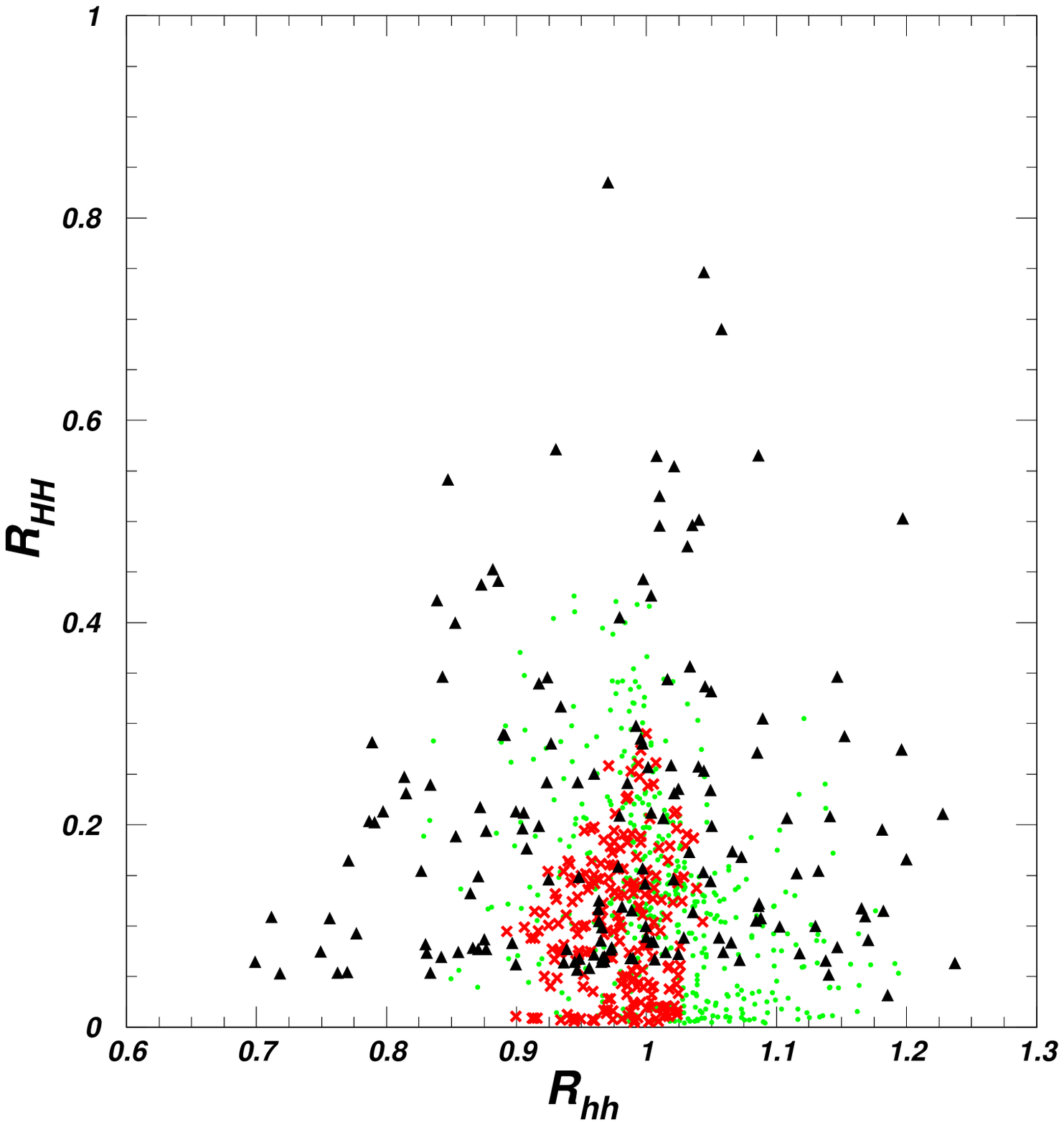,height=8.5cm}
 \epsfig{file=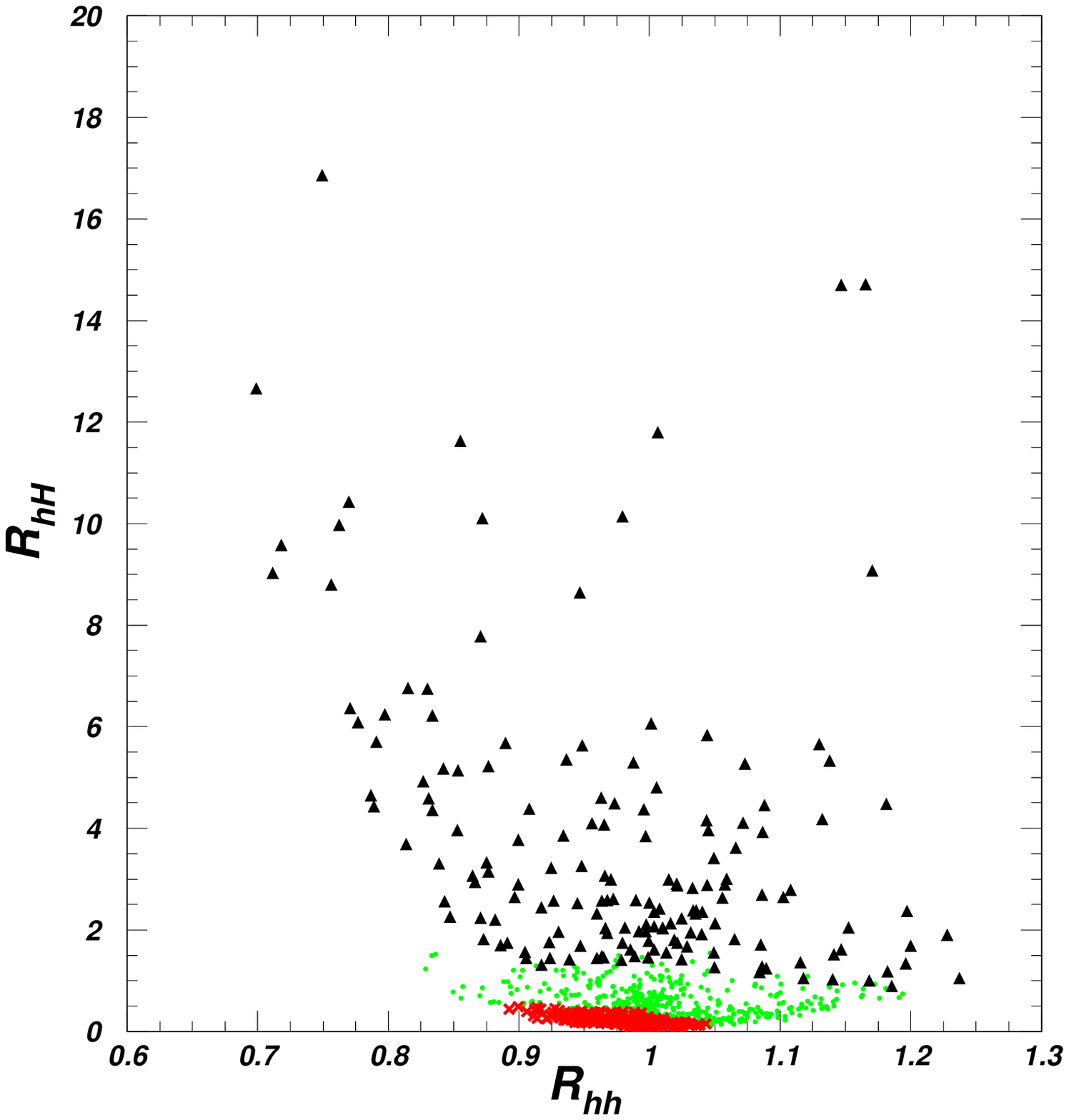,height=8.5cm}
\vspace{-0.3cm} \caption{Same as Fig. \ref{sbatb}, but projected on
the planes of $R_{hH}$ versus $R_{hh}$ and $R_{HH}$ versus $R_{hh}$.
Here the ratios denote the production rates via the gluon fusion
normalized to the SM cross section of $hh$ production.} \label{rsig}
\end{figure}

In Fig. \ref{sbatb}, we project the surviving samples on the planes
of $\cos(\beta-\alpha)$ versus $\tan\beta$ and $m_{12}^2$ versus
$\tan\beta$, respectively. At the 14 TeV LHC with an integrated
luminosity of 3000 $fb^{-1}$, the significance of SM is around
2$\sigma$ for the $b\bar{b}\gamma\gamma$ channel
\cite{yao,1502.00539}. Therefore, it should be difficult to probe
the $b\bar{b}\gamma\gamma$ channel of type-II 2HDM for
$R_{b\bar{b}\gamma\gamma}<2.0$. As shown in this figure,
$R_{b\bar{b}\gamma\gamma}>2.0$ favors $\tan\beta < 2$ ($\tan\beta <
1.2$ is excluded by $\Delta m_{B_s}$ and $\Delta m_{B_d}$),
$-1\times10^{5}~{\rm GeV^2}~<m_{12}^2<-3\times10^{4}~{\rm GeV^2}$
 and $0~{\rm GeV^2}<m_{12}^2<1.5\times10^{4}~{\rm GeV^2}$.
The various Higgs trilinear couplings are sensitive to $\tan\beta$ and
$m_{12}^2$. In addition, the top quark Yukawa couplings is
sensitive to $\tan\beta$, as shown in Eq. (\ref{yukawatop}).

To understand the allowed ranges of $R_{b\bar{b}\gamma\gamma}$, we
project the surviving samples on the planes of the Higgs couplings
in Fig. \ref{rhcoup}. The upper panel of Fig. \ref{rhcoup} shows
that the light CP-even Higgs trilinear coupling and its coupling to
top quark are restricted to be around the SM predictions,
respectively. The absolute value of the heavy CP-even Higgs coupling
to top quark is always suppressed, and allowed to be as low as 0.12
relative to the SM top quark Yukawa coupling. In some parameter space,
the absolute value of the Higgs trilinear couplings of $HHH$ and $HHh$
are respectively allowed to be as high as 15 and 10 relative to
the SM $hhh$ coupling. The absolute value of the coupling $Hhh$ is
always suppressed compared to the SM $hhh$ coupling due
to the suppression of $\cos(\beta-\alpha)$.

From Fig. \ref{rhcoup} we see that $R_{b\bar{b}\gamma\gamma}>2.0$
favors two different regions. In one region, the Higgs potential is
"decoupling", namely the  $hhh$ coupling is near the SM prediction
while other trilinear couplings of $HHH$, $HHh$ and $Hhh$ are very
small. Therefore, for the $gg\to hH$ production process, the
contributions of triangle diagrams will be sizably suppressed since
the couplings of $HHh$ and $Hhh$ are very small. This will sizably
soften the destructive interference between the triangle and box
diagrams, leading the enhancement of the cross section of $gg\to
hH$. In the other region, the coupling $HHh$ is much larger than the
SM $hhh$ coupling, which can make the contributions of the triangle
diagrams to overcome the box diagrams, and enhance the cross section
of $gg\to hH$. In addition, the upper-right panel of Fig.
\ref{rhcoup} shows that $R_{b\bar{b}\gamma\gamma}> 2.0$ favors the
absolute value of $y_t^H$ to be larger than 0.5, which avoids the
cross section of $gg\to hH$ to be sizably suppressed.

Although the couplings of $HHH$ and $HHh$ can be much larger than
the SM $hhh$ coupling, the cross section of $gg\to HH$ can not be
enhanced since there are destructive interference between the
triangle diagrams mediated by $H$ and $h$. Conversely, the cross
section of $gg\to HH$ is smaller than the SM cross section of $gg\to
hh$ since the $Ht\bar{t}$ coupling is suppressed. We show the cross
sections of $hh$, $hH$ and $HH$ in Fig. \ref{rsig}. This figure
shows that the cross section of $HH$ is smaller than 0.6 relative
the SM $hh$ prediction for most surviving samples. The cross section
of $hh$ is around the SM prediction, and the cross section of $hH$
can reach 17 times of the SM $hh$ prediction.

\subsection{Simulation results in a decoupling scenario}
As seen from the preceding section, the cross section of Higgs
pair production at the LHC can be sizably enhanced by
a large Higgs trilinear coupling in the 2HDM with two nearly degenerate
CP-even 125 GeV Higgs bosons, and as a result the Higgs pair signal is
observable at the LHC.
In the following we consider a "decoupling" scenario
in which the light CP-even Higgs has a SM-like cubic
self-coupling while other Higgs trilinear
couplings of the two CP-even Higgses are very small. In this
scenario, the $b\bar{b}\to hh,~hH,~HH$ processes can be neglected
since there is no enhancement of Higgs trilinear couplings.

We take a benchmark point
 \bea
&&\sin(\beta-\alpha)\simeq-0.999988,~\tan\beta\simeq1.232,~m_h=125.5
~{\rm{GeV}},~m_H\simeq126.0~{\rm{GeV}},\nonumber\\
&&m_A=595.65~{\rm{GeV}},~m_{H^\pm}=595.65~{\rm{GeV}},m_{12}^2=12304.0~\rm{GeV^2};\nonumber\\
&&y_{ht\bar{t}}=-0.996,~~y_{Ht\bar{t}}=0.82,~~y_{hb\bar{b}}=-1.006,~~y_{Hb\bar{b}}=-1.228,\nonumber\\
&&y_{hhh}=-0.99996,~~y_{HHH}=0.245,~~y_{HHh}=0.0598,~~y_{Hhh}=0.00552;\nonumber\\
&&Br(h\to \gamma\gamma)=1.969\times10^{-3},~~~~~Br(h\to
b\bar{b})=0.6119,\nonumber\\&&Br(H\to
\gamma\gamma)=9.702\times10^{-5},~~~~Br(H\to b\bar{b})=0.8385;\nonumber\\
&&\sigma(gg\to hh)_{LO}=16.74~fb,~~~~~\sigma(gg\to hH)_{LO}=50.4~fb.
\eea

\begin{figure}[tb]
 \epsfig{file=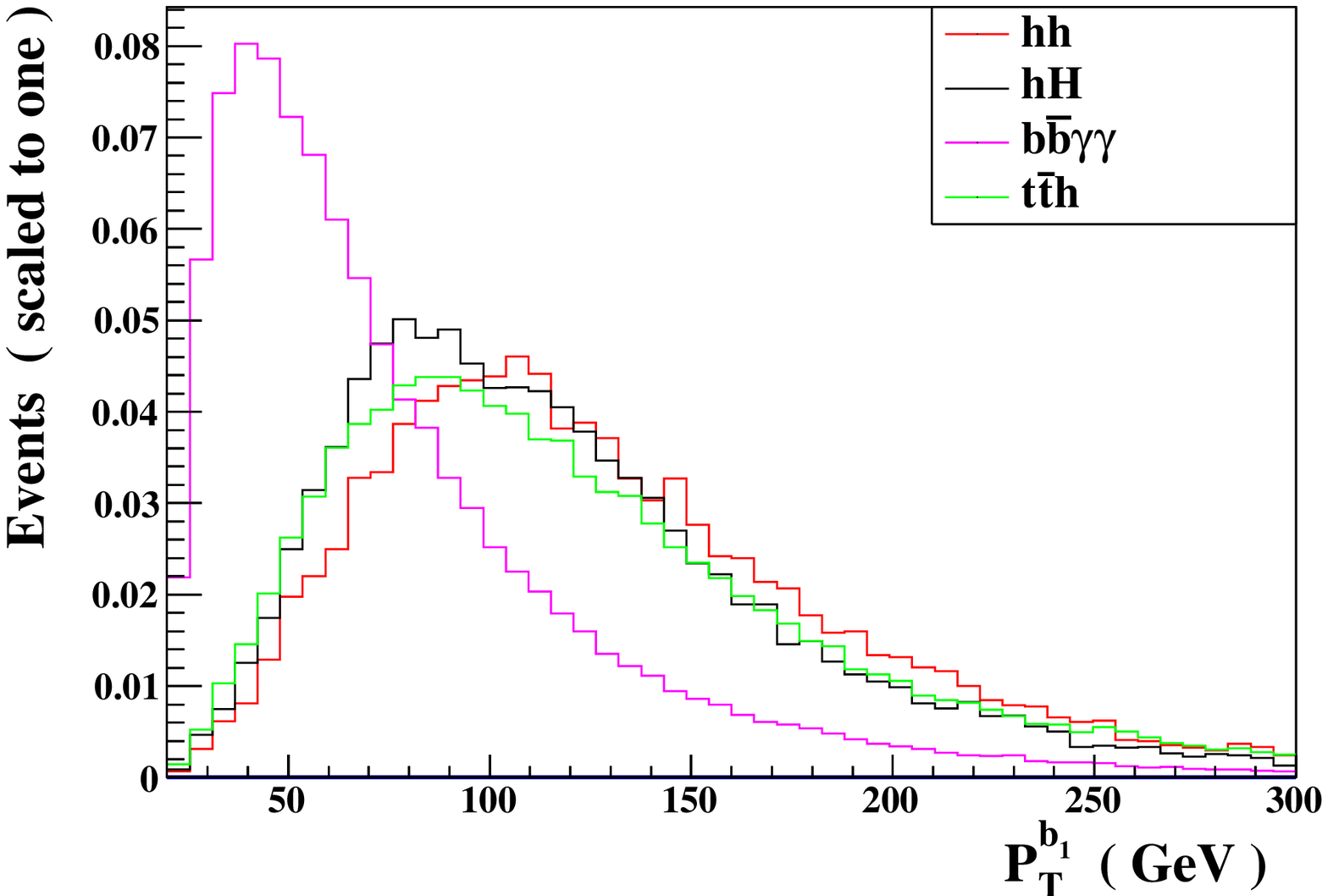,height=4.5cm}
 \epsfig{file=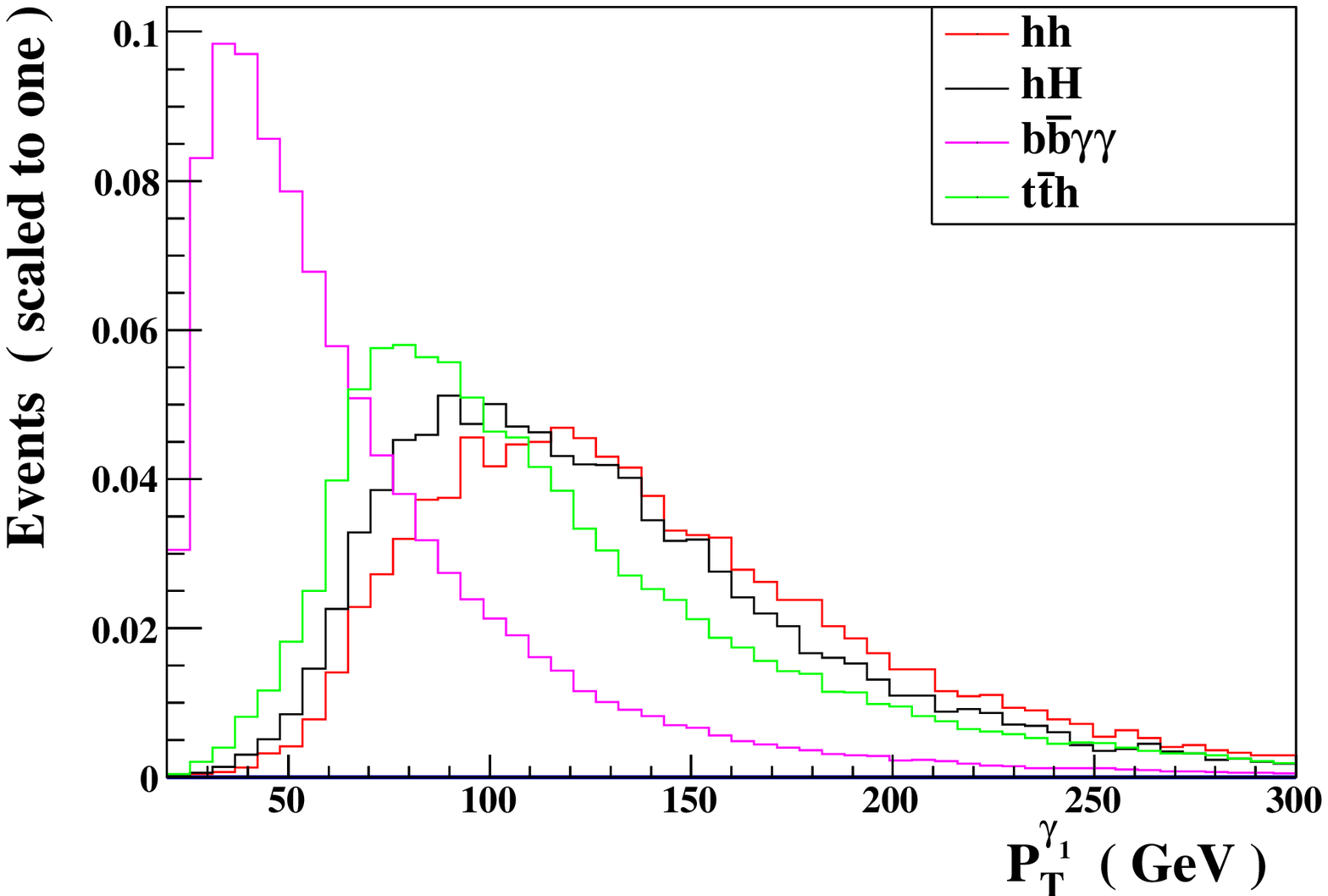,height=4.5cm}
   \epsfig{file=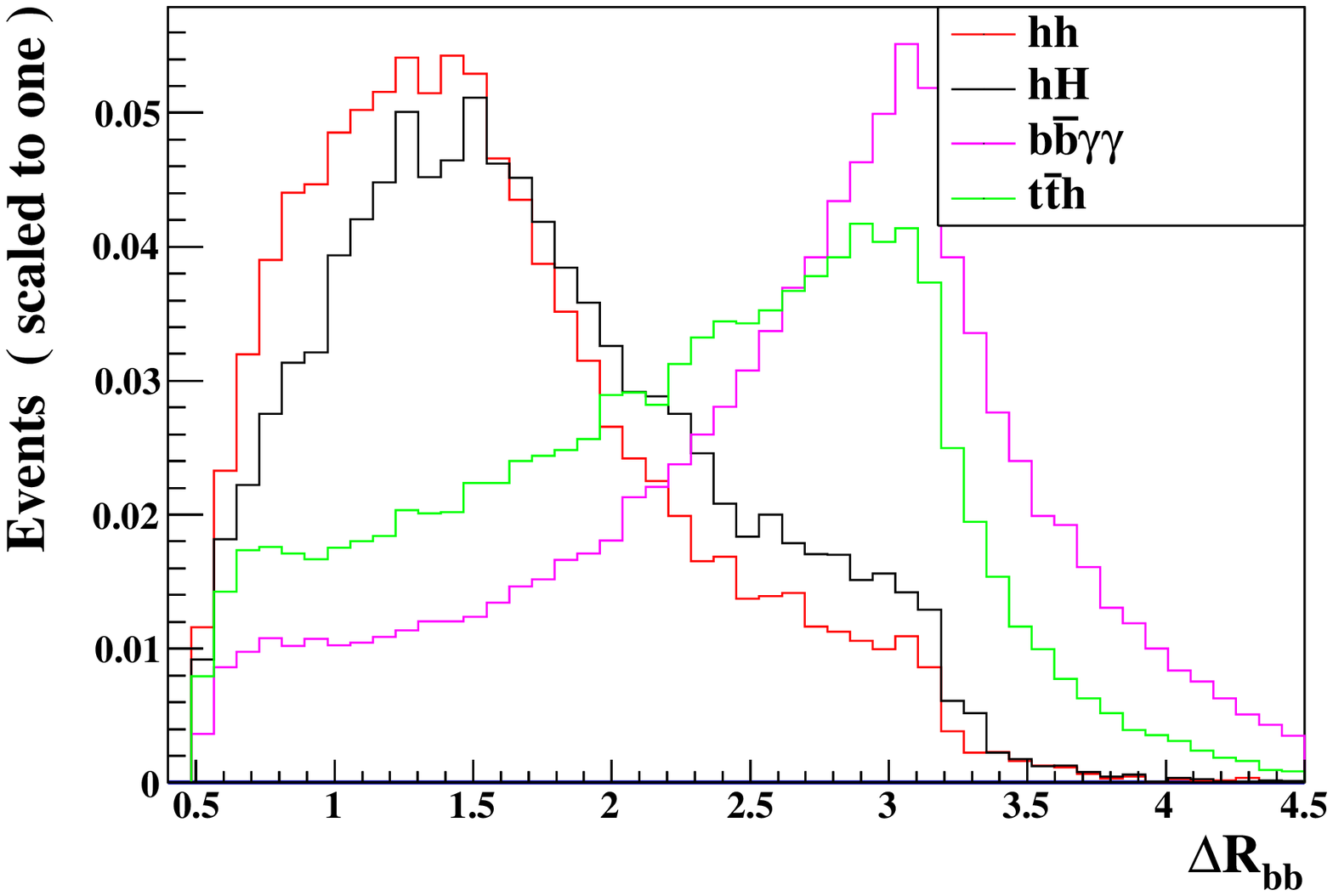,height=4.5cm}
  \epsfig{file=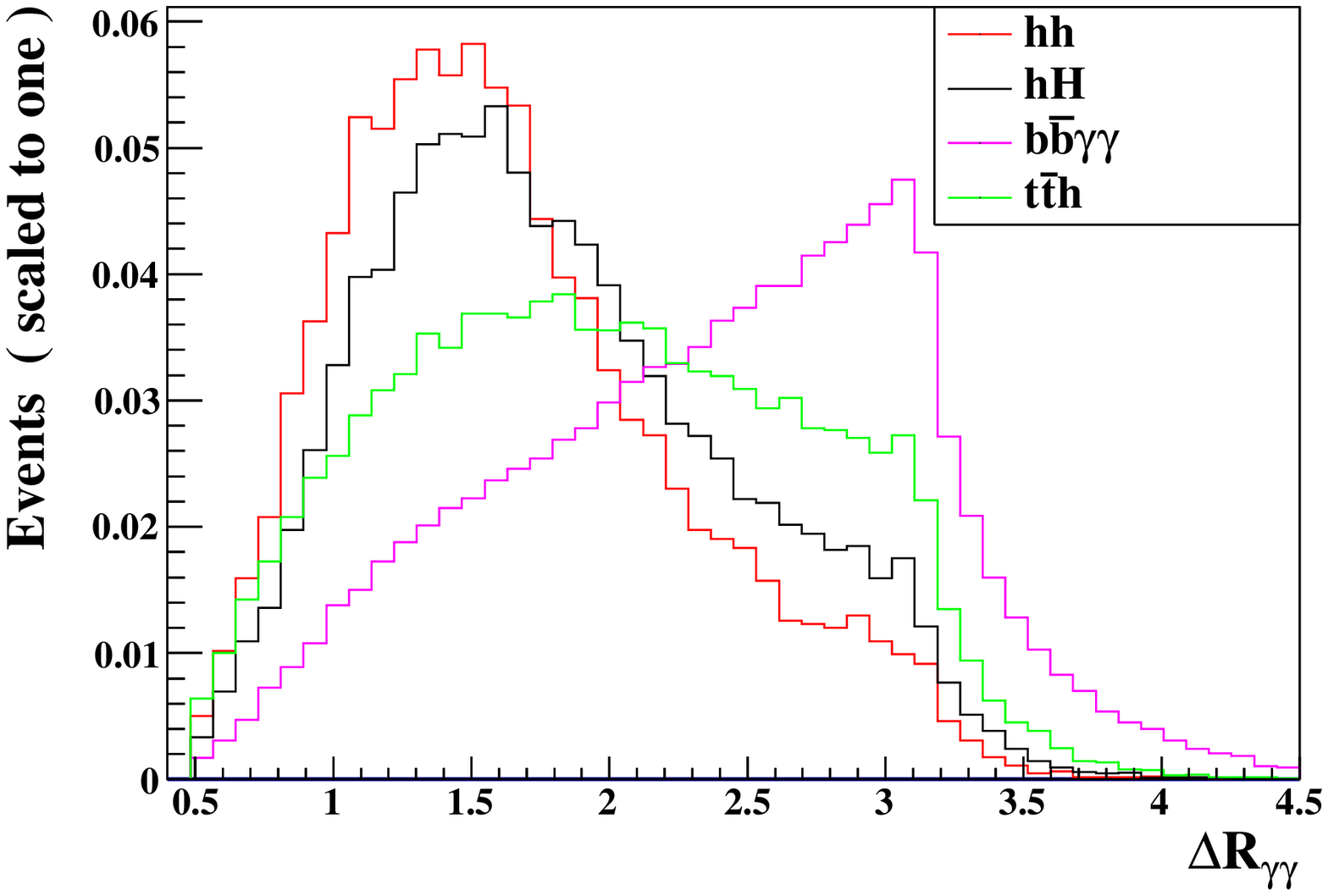,height=4.5cm}
   \epsfig{file=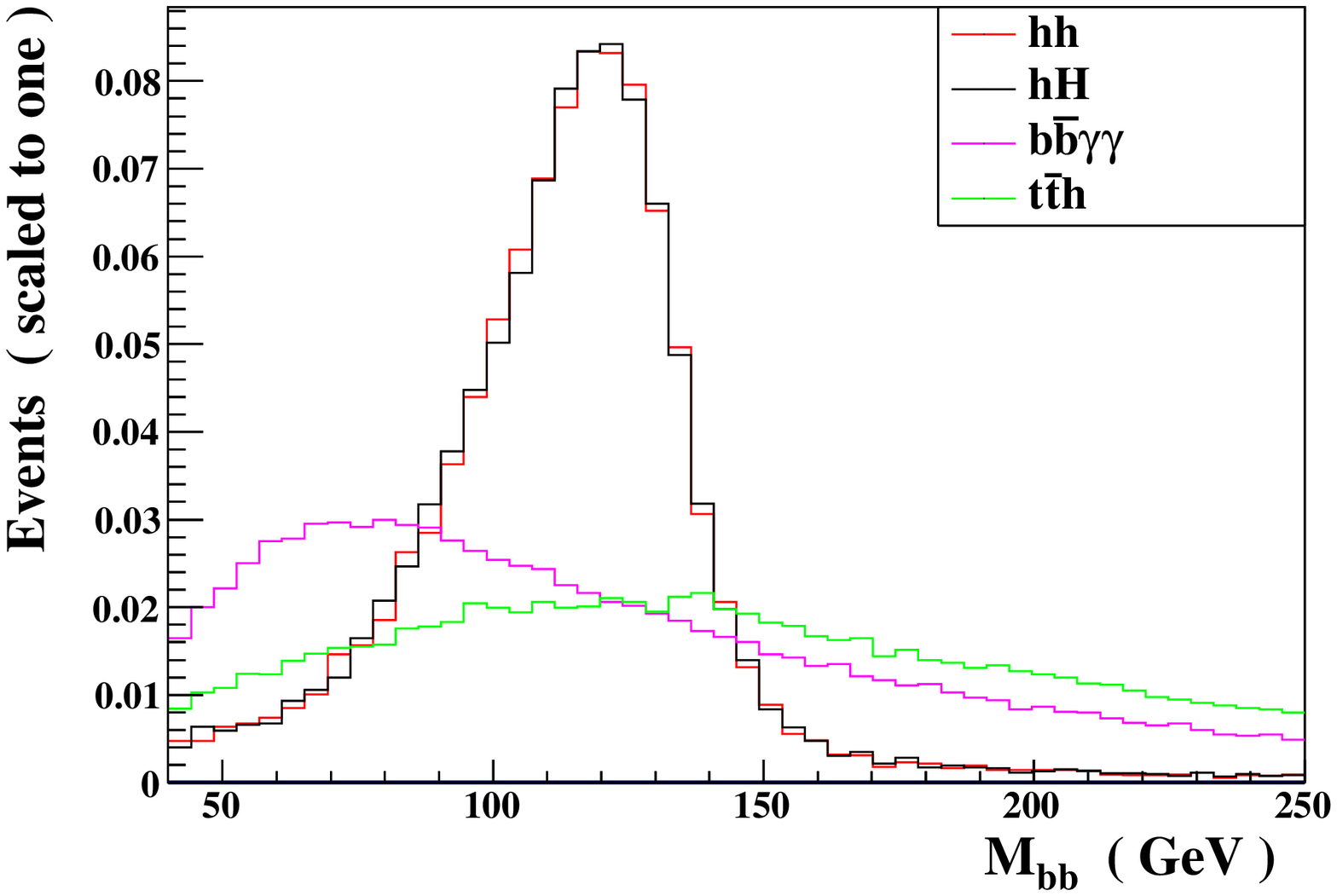,height=4.5cm}
  \epsfig{file=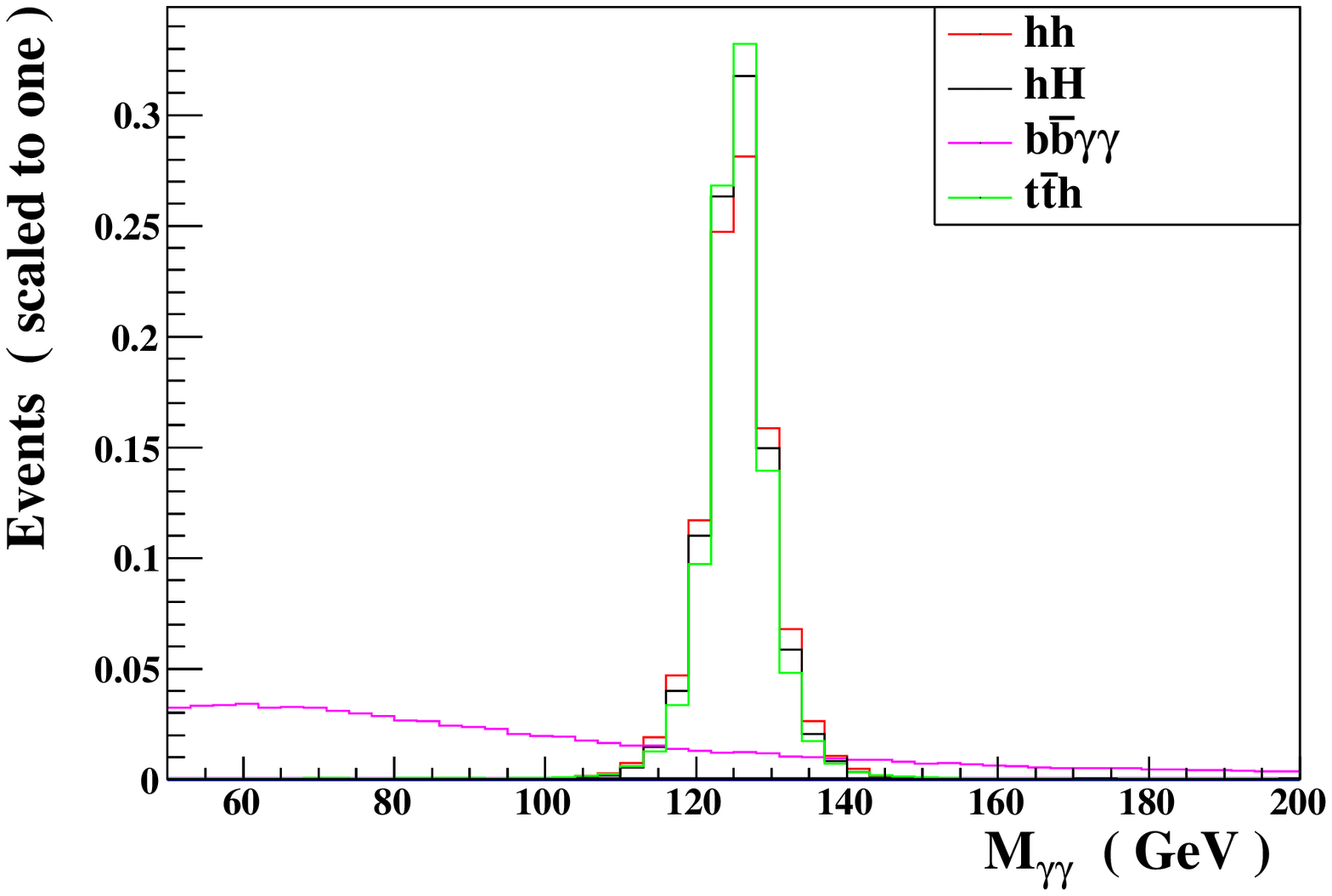,height=4.5cm}

\hspace*{-6.8cm}
    \epsfig{file=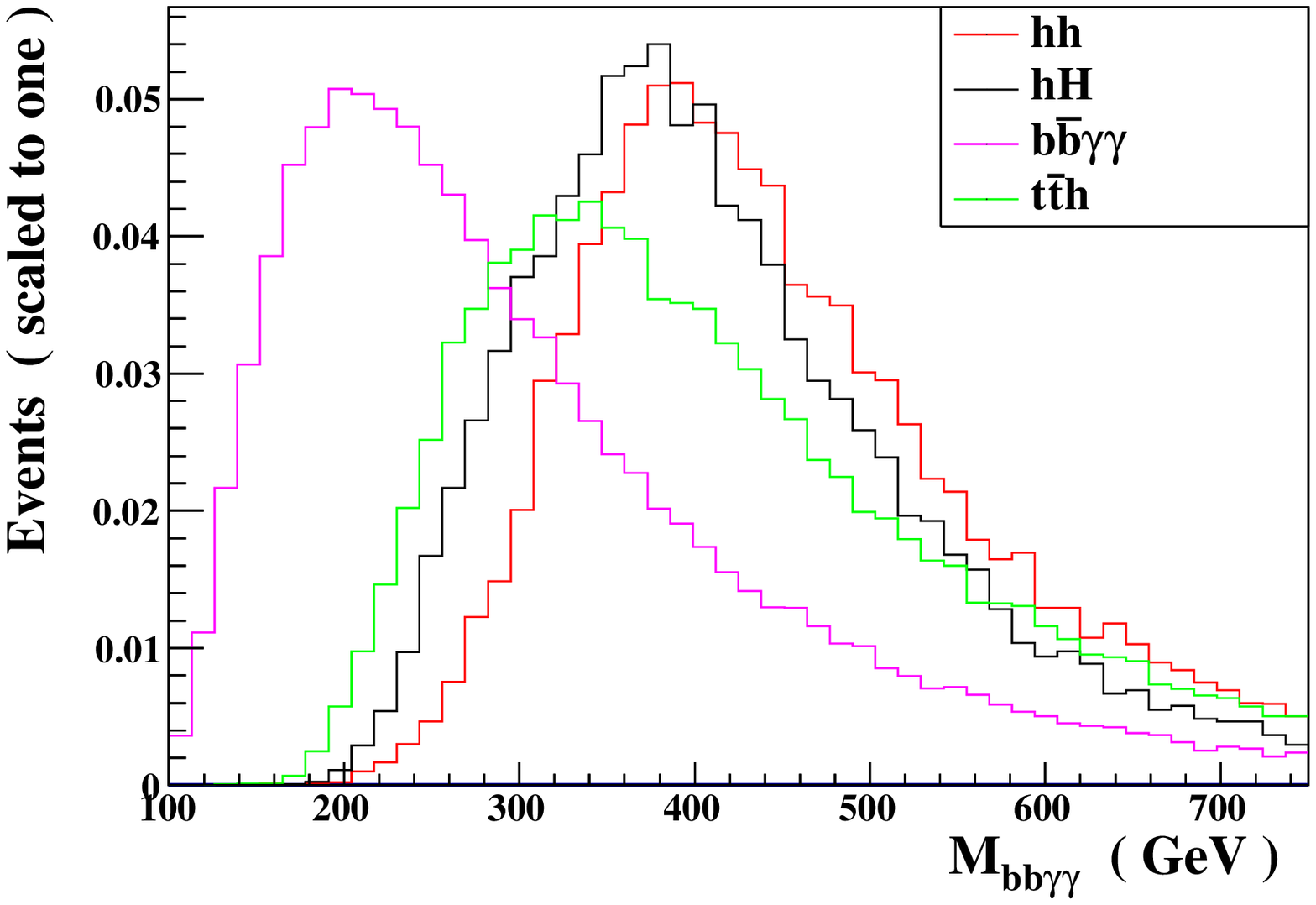,height=4.5cm}
\vspace{-0.6cm} \caption{The Higgs pair signal $b\bar{b}\gamma\gamma$ and
background distributions of $P_T^{b_1,\gamma_1}$, $\Delta
R_{bb,\gamma\gamma}$ and $M_{bb\gamma\gamma,bb,\gamma\gamma}$ at the
14 TeV LHC.} \label{distri}
\end{figure}

Since $Br(H\to \gamma\gamma)$ is very small, we neglect $g g \to H H
\to b\bar{b}\gamma\gamma$, and consider the $b\bar{b}\gamma\gamma$
signal from
\bea
&& g g \to h h \to b\bar{b}\gamma\gamma,\nonumber\\
&&g g \to h H \to b\bar{b}\gamma\gamma.
\eea
The main SM backgrounds include non-resonant $b \bar{b} \gamma\gamma$, $t
\bar{t} h$ $(t \bar{t}\to b\bar{b}+X,~h\to\gamma\gamma)$, $Zh$ $(Z
\to b \bar{b}, h\to\gamma\gamma)$ and $b\bar{b}h$
$(h\to\gamma\gamma)$. We neglect the subdominant reducible
backgrounds of $jj\gamma\gamma$ and $t\bar{t}\gamma\gamma$
\cite{hh2h1}. The QCD corrections are considered by including a $k$-factor, which is
2.27 for the signal \cite{hhk}, 2.0 for $b\bar{b}\gamma\gamma$
\cite{1502.00539}, 1.1 for $t \bar{t} h$ \cite{zhk}, 1.33 for $Zh$
\cite{zhk} and 1.2 for $b\bar{b}h$ \cite{bbhk}.

Fig. \ref{distri} shows the distributions of some kinematical
variables at the LHC with $\sqrt{s}=14$ TeV for the $hh$, $hH$,
$b\bar{b}\gamma\gamma$ and $t\bar{t}h$. The results of $Zh$ and $b\bar{b}h$ are not
shown since they are subdominant. According to the distribution
differences between the signal and backgrounds, we can improve the
ratio of signal to backgrounds by making some kinematical cuts.
First, we require the final states to include
two isolated photons and two $b$-jets, and further impose the
following cuts
\bea &&P_T^{b_1}
> 60~{\rm GeV},~~P_T^{b_2} > 25~{\rm GeV},~~P_T^{\gamma_1} > 60~{\rm
GeV},~~P_T^{\gamma_2} > 25~{\rm GeV},\nonumber\\
&&\Delta R_{bb} > 0.4,~~\Delta R_{\gamma\gamma} > 0.4,~~\Delta
R_{b\gamma} > 0.4,\nonumber\\
&&|\eta_b|<2.5,~~~~~~|\eta_\gamma|<2.5,\nonumber\\
&&M_{bb}> 30~{\rm GeV},~~M_{\gamma\gamma}> 30~{\rm
GeV},~~M_{bb\gamma\gamma}> 350~{\rm GeV},\label{basiccut}
\eea
where $P_T^{b_1}$ and $P_T^{\gamma_1}$ denote the transverse momentum of
the hardest $b$-jet and photon, and $P_T^{b_2}$ and $P_T^{\gamma_2}$
for the second hardest $b$-jet and photon. $\Delta R = \sqrt{(\Delta
\phi)^2+(\Delta \eta)^2}$ is the particle separation with $\Delta
\phi$ and $\Delta \eta$ being the separation in the azimuthal angle
and rapidity respectively. The cuts of the invariant mass of two
$b$-jets and two photon $M_{bb\gamma\gamma}$, $P_T^{b_1}$ and
$P_T^{\gamma_1}$ can suppress the backgrounds sizably, especially
the largest background $b\bar{b}\gamma\gamma$.

The photon pair is further restricted to have
\beq
\Delta R_{\gamma\gamma} < 2.0,~~~115~{\rm GeV} < M_{\gamma\gamma}<135~{\rm GeV}.\label{cutphoton}
\eeq
The $b$-quark pair is restricted to have
\beq
\Delta R_{bb} < 2.0,~~~100~{\rm GeV} < M_{bb}<140~{\rm GeV}.\label{cutb}
\eeq
Since the two photons (two $b$ quarks) in the
signals are from the Higgs decay, the signal rates peak
at their invariant mass around the Higgs mass with relative
small separation. The cuts in Eqs. (\ref{cutphoton}) and (\ref{cutb})
play the dominant role in suppressing the backgrounds.

Finally, we make some cuts which can specially suppress the background
$t\bar{t}h$. Since $W^{\pm}$ will decay into the charged leptons and
jets, the background $t\bar{t}h$ tends to include additional charged
leptons and more jets. Therefore, we will veto the following case
\beq P_T^\ell
> 20~{\rm GeV}~~{\rm or}~~P_T^{j_8}> 20~{\rm GeV},\label{cuttth}
\eeq
where $P_T^{j_8}$ denotes the transverse momentum of the 8th hardest
jet.

The resulting cut flow is shown in Table. \ref{cutflow}. The
$b\bar{b}\gamma\gamma$ and $t\bar{t}h$ are the two major
backgrounds. After imposing the above cuts, the events from signal
$hH$ are approximately 1.6 times of those of $hh$. Since $h$ and $H$
are the degenerate 125.5 GeV Higgses, the total signal events are
from $hh$ and $hH$, whose significance can reach 5$\sigma$ at the 14
TeV LHC with an integrated luminosity of 3000 $fb^{-1}$. If there is
sizable mass splitting between $h$ and $H$, the cuts in Eq.
(\ref{cutphoton}) and Eq. (\ref{cutb}) will hurt the events of $hH$
inevitably and hence suppress the significance. Therefore, the
degeneracy between $h$ and $H$ plays the key role in enhancing the
significance for such a "decoupling" scenario.

\begin{table}
 \caption{The cut flow for the signal and background event numbers at the 14 TeV LHC with
 an integrated luminosity of 3000 fb$^{-1}$ for the $b\bar{b}\gamma\gamma$
 channel. The two culumns labeled '$hh$' and '$hH$' are for the Higgs pair signal
while other culumns are for the backgrounds.}\vspace{0.5cm}
  \setlength{\tabcolsep}{2pt}
  \centering
  \begin{tabular}{|c|c|c|c|c|c|c|c|}
    \hline
     $\sqrt{s}=14~{\rm TeV}$,~3 $ab^{-1}$&~~$hh$~~& ~~$hH$~~ &~~$b\bar{b}\gamma\gamma$~~&~~$t\bar{t}h$~~&~~$Zh$
     ~~&~~$b\bar{b}h$~~&~~$S/\sqrt{B}$~~\\
    \hline
     ~~${\rm after~cut~in~Eq.(\ref{basiccut})}$~~
     & $38.4$ & $63.3$
     & $15999$ & $246.8$& $22.5$& $9.8$& $0.8$
     \\
     \hline
     ~~${\rm after~cut~in~Eq.(\ref{cutphoton})}$~~
      & $29.9$& $48.1$
      & $679.5$ & $152.5$& $18.4$& $4.7$& $2.7$
     \\
     \hline
         ~~${\rm after~cut~in~Eq.(\ref{cutb})}$~~
      & $18.7$ & $29.9$
      & $74$ & $16.8$& $2.7$& $0.4$& $5.1$
     \\
     \hline
         ~~${\rm after~cut~in~Eq.(\ref{cuttth})}$~~
      & $18.6$ & $29.7$
      & $74$ & $10.4$& $2.7$& $0.4$& $5.2$
     \\
     \hline
      \end{tabular}
\label{cutflow}
\end{table}

\section{Conclusion}
In this work we discussed a special scenario in the type-II 2HDM
where the $b\bar{b}\gamma\gamma$ channel of the Higgs pair
production can be enhanced due to the two nearly degenerate 125 GeV
Higgses. We considered various theoretical and experimental
constraints and found that in the allowed parameter space some
trilinear Higgs couplings can be larger than the SM value by an
order and the signal $b\bar{b}\gamma\gamma$ can be sizably enhanced.
We also considered  a "decoupling" scenario where the light CP-even
Higgs has the SM-like cubic self-coupling while other trilinear
Higgs couplings are very small. From a detailed simulation on the
signal $b\bar{b}\gamma\gamma$ and backgrounds, we found that the
$hh$ and $hH$ production channels can jointly enhance the
statistical significance to 5$\sigma$ at the 14 TeV LHC with an
integrated luminosity of 3000 fb$^{-1}$. Therefore, the degenerate
$h$ and $H$ play the vital role in enhancing the significance for
probing the Higgs potential "decoupling" scenario.

\section*{Acknowledgment}
We would like to thank Tao Liu, Olivier Mattelaer, Lei Wu and Shuo
Yang for helpful discussions, and Antonio Pich for reading of the
manuscript and useful suggestions. This work has been supported in
part by the Spanish Government and ERDF funds from the EU Commission
[Grant No. FPA2011-23778], by the Spanish {\it Centro de Excelencia
Severo Ochoa} Programme [Grant SEV-2014-0398], and by the National
Natural Science Foundation of China under grant No. 11275245,
10821504 and 11135003.


\begin{thebibliography}{99}
\bibitem{cmsh} S. Chatrchyan et al. [CMS Collaboration], \PLB716, 30 (2012).

\bibitem{atlh} G. Aad et al. [ATLAS Collaboration], \PLB716, 1 (2012).

\bibitem{bbtautau} V. Barger, L. L. Everett, C. B. Jacksonand, G. Shaughnessy, \PLB728, 433 (2014).

\bibitem{subjet} D. E. FerreiradeLima, A. Papaefstathiou, M. Spannowsky, \JHEP08, 030 (2014);
A. Papaefstathiou, L. L. Yang, J. Zurita, \PRD87, 011301 (2013);
A. J. Barr, M. J. Dolan, C. Englertand, M. Spannowsky, \PLB728, 308 (2014);
D. E. Ferreira de Lima, A. Papaefstathiou, M. Spannowsky, \JHEP1408, 030 (2014);
C. Englert, F. Krauss, M. Spannowskyand, J. Thompson, \PLB743, 93 (2015);
Z. Kang, P. Ko, J. Li, arXiv:1504.04128.

\bibitem{yao} W. Yao, arXiv:1308.6302.

\bibitem{1502.00539} A. Azatov, R. Contino, G. Panico, M. Son, arXiv:1502.00539.

\bibitem{caohh} Q.-H. Cao, B. Yan, D.-M. Zhang, H. Zhang, arXiv:1508.06512.

\bibitem{hh2h1} V. Barger, L. L. Everett, C. B. Jackson, A. D. Peterson, G.
Shaughnessy, \PRD90, 095006 (2014).

\bibitem{hh2h2} B. Hespel, D. Lopez-Val, E. Vryonidou, \JHEP1409, 124 (2014);
U. Baglio, O. Eberhardt, U. Nierste, M. Wiebusch, \PRD90, 015008
(2014); L.-C. Lu, C. Du, Y. Fang, H.-J. He, H. Zhang,
arXiv:1507.02644.

\bibitem{100tev} J. Baglio, {\it et al.}, \JHEP1304, 151 (2013);
J. Baglio, arXiv:1408.6066; A. J. Barr,  {\it et al.}, \JHEP1502,
016 (2015); Q. Li, Z. Li, Q.-S. Yan, X. Zhao, \PRD92, 014015 (2015);
I. Hinchliffe, {\it et al.}, arXiv:1504.06108; H.-J. He, J. Ren, W.
Yao, arXiv:1506.03302; Z.-L. Han, R. Ding, Y. Liao,
arXiv:1506.08996; C.-T. Lu, J. Chang, K. Cheung, J. S. Lee,
arXiv:1505.00957; A. Papaefstathiou, arXiv:1504.04621; M. Slawinska,
W. v. den Wollenberg, B. v. Eijk, S. Bentvelsen, arXiv:1408.5010.


\bibitem{hhlh} L. Wang, W. Wang, J. M. Yang, H. Zhang, \PRD76, 017702 (2007);
J.-J. Liu, {\it et al.}, \PRD70, 015001 (2004);
C. O. Dib, R. Rosenfeld, A. Zerwekh, \JHEP05, 074 (2006).


\bibitem{hhsusy} U. Ellwanger, arXiv:1306.5541;
C. Han, {\it et al.}, \JHEP1404, 003 (2014);
J. Cao, L. Shang, P. Wan, J. M. Yang, \JHEP1304, 134 (2013);
S. Dawson, C. Kao, Y. Wang, \PRD77, 113005 (2008).


\bibitem{type-ii} L. J. Hall, M. B. Wise, \NPB187, 397 (1981);
 J. F. Donoghue, L. F. Li, \PRD19, 945 (1979).

\bibitem{gunion} J. F. Gunion, Y. Jiang, S. Kraml, \PRL110, 051801 (2013);
P. M. Ferreira, R. Santos, H. E. Haber, J. P. Silva, \PRD87, 055009
(2013); Y. Grossman, Z. Surujon, J. Zupan, \JHEP1303, 176 (2013); M.
Chabab, M. C. Peyranere, L. Rahili, \PRD90, 035026 (2014); A. David,
J. Heikkila, G. Petrucciani, \EPJC75, 49 (2015).


\bibitem{2h-poten} R. A. Battye, G. D. Brawn, A. Pilaftsis, \JHEP1108, 020 (2011).

\bibitem{a2hm-1} A. Pich, P. Tuzon, \PRD80, 091702 (2009).

\bibitem{2hc-1} D. Eriksson, J. Rathsman, O. St{\aa}l, \CPC181, 189 (2010); \CPC181, 833 (2010).

\bibitem{spriso} F. Mahmoudi, \CPC180, 1579-1673 (2009).

\bibitem{hb} P. Bechtle, {\it et al.}, \CPC181, 138 (2010); \EPJC74, 2693 (2014).

\bibitem{pdg2014} K. A. Olive etal. [Particle Data Group], \CHC38, 090001 (2014).

\bibitem{nloct} C. Degrande, arXiv:1406.3030.

\bibitem{fr} A. Alloul, {\it et al.}, \CPC185, 2250 (2014).

\bibitem{ufo} C. Degrande, {\it et al.}, \CPC183, 1201 (2012).

\bibitem{mg5} J. Alwall,  {\it et al.}, \JHEP1407, 079 (2014);
V. Hirschi, O. Mattelaer, arXiv:1507.00020.

\bibitem{pythia} T. Sjostrand, S. Mrenna, P. Z. Skands, \JHEP0605, 026 (2006).

\bibitem{delphes} J. de Favereau, {\it et al.}, \JHEP1402, 057 (2014).

\bibitem{ma5} E. Conte, B. Fuksand, G. Serret, \CPC184, 222 (2013).

\bibitem{chi} J. R. Espinosa, C. Grojean, M. Muhlleitner, M. Trott, \JHEP1205, 097 (2012);
G. Belanger, B. Dumont, U. Ellwanger, J. F. Gunion, S. Kraml, \JHEP1302, 053 (2013);
P. P. Giardino, K. Kannike, M. Raidal, A. Strumia, \JHEP1206, 117 (2012);
B. Dumont, S. Fichet, G. Gersdorff, \JHEP1307, 065 (2013);
J. Cao,  {\it et al.}, \JHEP1203, 086 (2012);
J. S. Lee, P. Y. Tseng, \JHEP1305, 134 (2013).

\bibitem{higgsdata} G. Aad et al. [ATLAS Collaboration], \PRD90, 112015 (2014); \PRD90, 052004;
S. Chatrchyan et al. [CMS Collaboration], \EPJC74, 3076 (2014); \PRD89, 092007 (2014);
\JHEP1401, 096 (2014); \PRD89, 012003 (2014); \JHEP1405, 104 (2014);
Talk by K. Herner, "Studies of the Higgs boson properties at D0", ICHEP 2014,
Spain.

\bibitem{kmdata} K. Cheung, J. S. Lee, P.-Y. Tseng, \PRD90, 095009 (2014).

\bibitem{hhk} D. deFlorian, J.Mazzitelli, \PRL111, 201801 (2013).

\bibitem{zhk} S. Dittmaier {\it et al.} [LHC Higgs Cross Section group], arXiv:1101.0593.

\bibitem{bbhk} S. Dawson, C. B. Jackson, L. Reina, D. Wackeroth, \PRL94, 031802 (2005).

\end{thebibliography}
\end{document}